\documentclass[10pt]{article}
\usepackage{amsmath}
\usepackage{amssymb}
\usepackage{graphics}
\usepackage{graphicx}
\usepackage{natbib}

\begin{document}
\bibliographystyle{plainnat}
\setcitestyle{numbers,square}
\title{JET BENDING AND VELOCITY PROFILES}

\author{I. Liodakis$^1$, V. S. Geroyannis$^2$, V. G. Karageorgopoulos$^3$ \\
        $^{1}$Department of Physics, University of Crete, Greece          \\
        $^{2,3}$Department of Physics, University of Patras, Greece       \\ \\
        $^1$liodakis@physics.uoc.gr, $^2$vgeroyan@upatras.gr,             \\
        $^3$vkarageo@upatras.gr}

\maketitle

\begin{abstract}
In this study, we consider the problem of geometry of curved jets. We work with a simplified two-dimensional kinematical model assuming that the jet velocity is varying. We prescribe the velocity variation and study two cases concerning parsec and kilo-parsec scale jets by applying some simplified numerical simulations. For parsec scale jets, we use some general expressions and derive the shape of jets for certain numerical cases. For kilo-parsec scale jets, we study the cases of 4C53.37 and 1159+583 of Abell 2220 and Abell 1446 galaxy clusters, respectively, in an effort to test our model and compare our numerical results with observational data.  
\end{abstract}

\section{Introduction}
Jets are an interesting issue in astrophysics. We know how they are produced and propagate through space, but there are still some questions regarding their interaction with their environment. In theory, galactic jets seem to be streams of plasma directed perpendicular to an accretion disk surrounding a supermassive  black hole. Observations reveal curved jets. Such jet bending is due to the interaction of the jet with a nebula, a gas cloud, a density anisotropy, or any other cause that can be expressed as a pressure gradient while the jet is journeying through the galactic center or the intergalactic medium.

The bending of a jet depends on its velocity and on the external pressure gradient (\citep{FH}, Secs.~I and III). Assuming that (i) the pressure gradient is distance-dependent (\citep{HU}, Fig.~5.4),
\begin{equation}
P = \, \frac{P_0}{1 + y^2/y^2_0} \, ,
\end{equation} 
where $P_0$ and $y_0$ are model parameters, and that (ii) this behaviour is the same for any jet, then the difference in jet bending is due to the velocity profile. 

Even though small scale jets are usually considered to be accelerating to speeds near the speed of light, velocities for parsec and kilo-parsec scales are of order $\sim 10^3 \, \mathrm{km \, s^{-1}}$ (cf. \citep{PS}, Table~1);
and they are generally considered constant. 
For galactic jets owing to AGN, which can reach lengths of up to a few mega-parsecs, it is probably of lower significance to consider time-varying velocities. Still, given the scales involved, we cannot assume that they are constant, especially when their interaction with the pressure gradient is taken into account.
Accordingly, we have to examine two scenarios: either an accelerating jet, until it becomes relativistic or destroyed after interacting with an external pressure, or a decelerating one.

\section{Jet Model}
For our jet model, we adopt a two-dimensional kinematical model (\citep{HU}, Eq.~5.17; we use here the definitions and symbols of this investigation),
\begin{equation}
\frac{d^2 f}{d x^2} \, \left[ 1+ \left( 
\frac{d f}{d x} \right)^2 \right]^{-3/2} = \,
\frac{1}{\rho u^2} \, \, \frac{\partial P}{\partial R}|_{\perp},
\end{equation}
where (\citep{HU}, Eq.~5.18)
\begin{equation}
\frac{\partial P}{\partial R}|_{\perp} = 
\sin(a) \, \, \frac{\partial P}{\partial x} \, - \, 
              \cos(a) \, \, \frac{\partial P}{\partial y},
\end{equation}
with (\citep{HU}, Eq.~5.19)
\begin{equation}
\tan(a) = \, \frac{d f}{d x}.
\end{equation}
Assuming that the pressure gradient is a function of the $y$-coordinate, we obtain the dimensionless form (\citep{HU}, Eq.~5.22)  
\begin{equation}
x = x_0 + \int^y_0 \left[L \, e^{-2P(f) / \rho \, u^2} - 1 \right]^{-1/2} \, df,
\end{equation}
where $x_0$ is the integration constant and $L$ a parameter associated with the angle the jet meets the pressure gradient (\citep{HU}, Eq.~5.23),
\begin{equation}
\tan(i) = \left( L - 1 \right)^{-1/2}.
\end{equation}
There is a critical value $L_\mathrm{crit}$ (\citep{HU}, remarks following Eq.~(5.27) and Fig.~5.5), such that, when below that value, the jet stops propagating forward and returns towards its source.

\section{The Computations}
To compile our code, we use the gfortran compiler, licensed under the GNU General Public License (GPL; http://www.gnu.org/licenses/gpl.html). gfortran is the name of the GNU Fortran compiler belonging to the GNU Compiler Collection (GCC; http://gcc.gnu.org/). In our computer, it has been installed by the TDM-GCC ``Compiler Suite for Windows'' (http://tdm-gcc. tdragon.net/), which is free software distributed under the terms of the GPL.
Subroutines required for standard numerical procedures are taken from the SLATEC Common Mathematical Library, which is an extensive public-domain Fortran Source Code Library, incorporating several public-domain packages. The full SLATEC release is available in http://netlib.org/slatec/. In particular, to compute integrals we use the subroutine \texttt{DGAUS8}, which integrates real functions in one variable along finite intervals on the basis of an adaptive 8-point Legendre-Gauss algorithm, and which is appropriate for high accuracy integrations.

\section{Parsec and Kilo-parsec Scale Jets}
\begin{figure}
\centering
\includegraphics[scale=0.5]{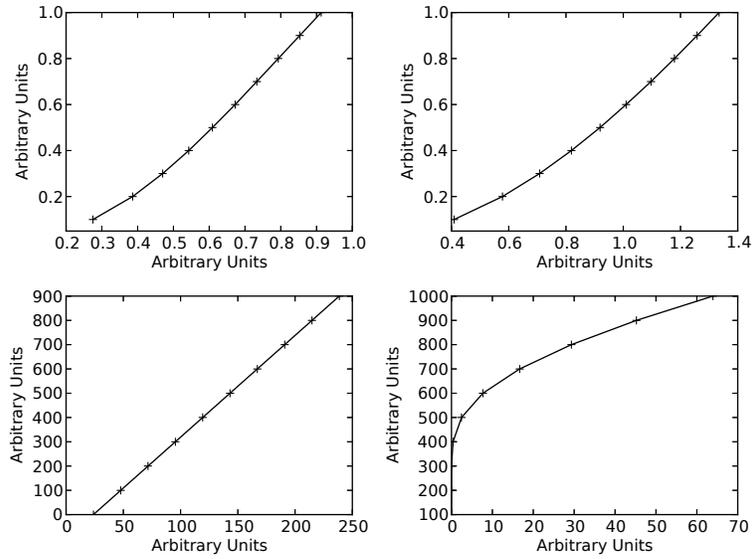}   
\caption{Velocity profiles corresponding to four cases. Top left: accelerating exponentially. Top right: accelerating linearly. Bottom left: decelerating exponentially. Bottom right: decelerating linearly. \label{f1}}
\end{figure}

After some preliminary tests of the code, we examine four different cases for the velocity profile. Two cases have to do with an accelerating jet (linear and nonlinear), and further two with a decelerating one (linear and nonlinear). For the nonlinear case we adopt a simplified exponential behaviour,
\begin{equation}
u = u_0 \, \mathrm{exp}(\pm y).
\end{equation}
We examine several velocity profiles and scales for all cases and show their graphs in Fig.~\ref{f1}. Regarding the form we expect to have for this model, it is clear that the jets must be decelerating linearly, a conclusion also derived in \citep{DY} (Sec.~3.3). With that result in mind, we proceed in examining the cases for parsec and kilo-parsec scale jets.

\subsection{Parsec scale jets} 
In the case of parsec scale jets, we assume that the bending is caused by an external pressure gradient interacting with the jet while moving through the galactic center. We examine sixteen different sets of parameters. In detail, we introduce four ``velocity coefficients'', $\lambda_1 = 0.1$, $\lambda_2 = 0.4$, $ \lambda_3 = 0.7$, $\lambda_4 = 1.0$, in order for the velocity to vary from one to two orders of magnitude (\citep{DY}, Sec.~4)
\begin{equation}
u_i = u_0 - \lambda_i \, y
\end{equation}
In addition, for each velocity coefficient $\lambda_i$, we use  four values of the parameter $L$, $L_1 = 2.718$, $L_2 = 4$, $L_3 = 6.8482$, $L_4 = 10.475$, corresponding to angles $i_1 = \pi/2$, $i_2 = \pi/4$, $i_3 = \pi/8$, and $i_4 = \pi/10$, respectively. Numerical experiments with several angles show that these particular values lead to distinguishable graphs (since close angles yield almost identical curves).

\begin{figure}
\centering
\includegraphics[scale=0.5]{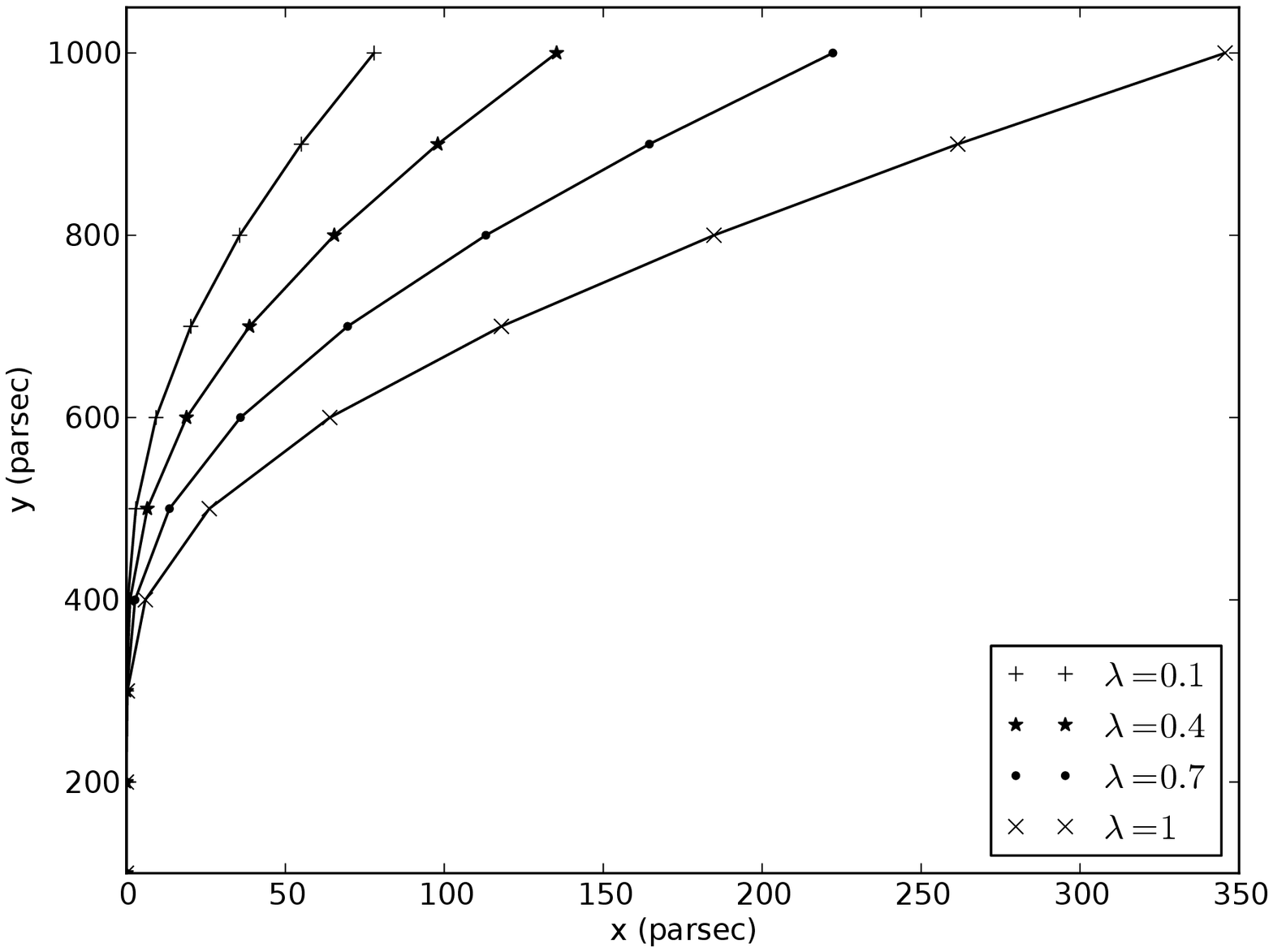} 
\caption{Jet paths with interaction constant $L=2.718$ for all velocity coefficients $\lambda_i$. \label{f2}}
\end{figure}

\begin{figure}
\centering
\includegraphics[scale=0.5]{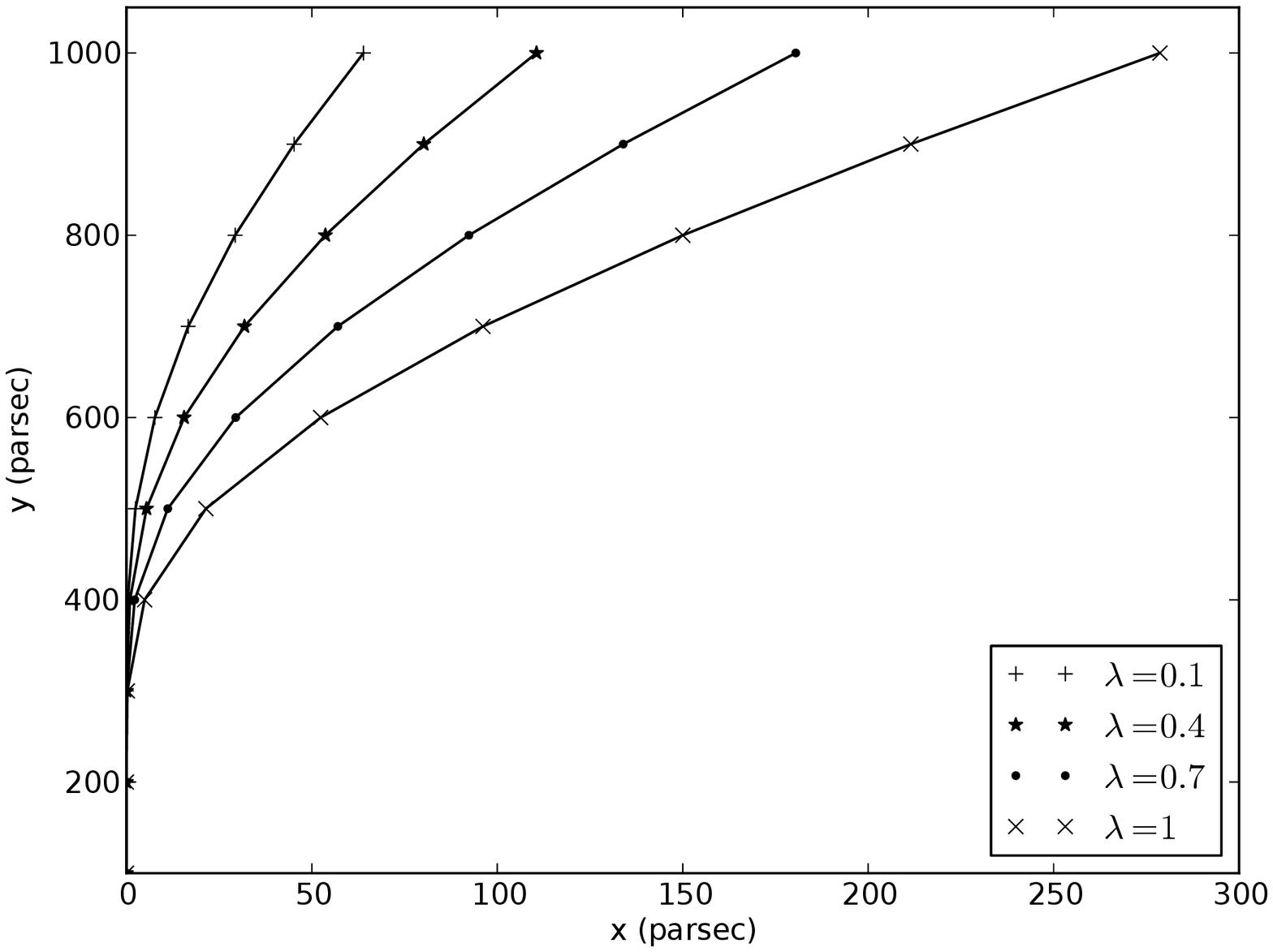} 
\caption{Jet paths with interaction constant $L=4$ for all velocity coefficients $\lambda_i$. \label{f3}}
\end{figure}

\begin{figure}
\centering
\includegraphics[scale=0.5]{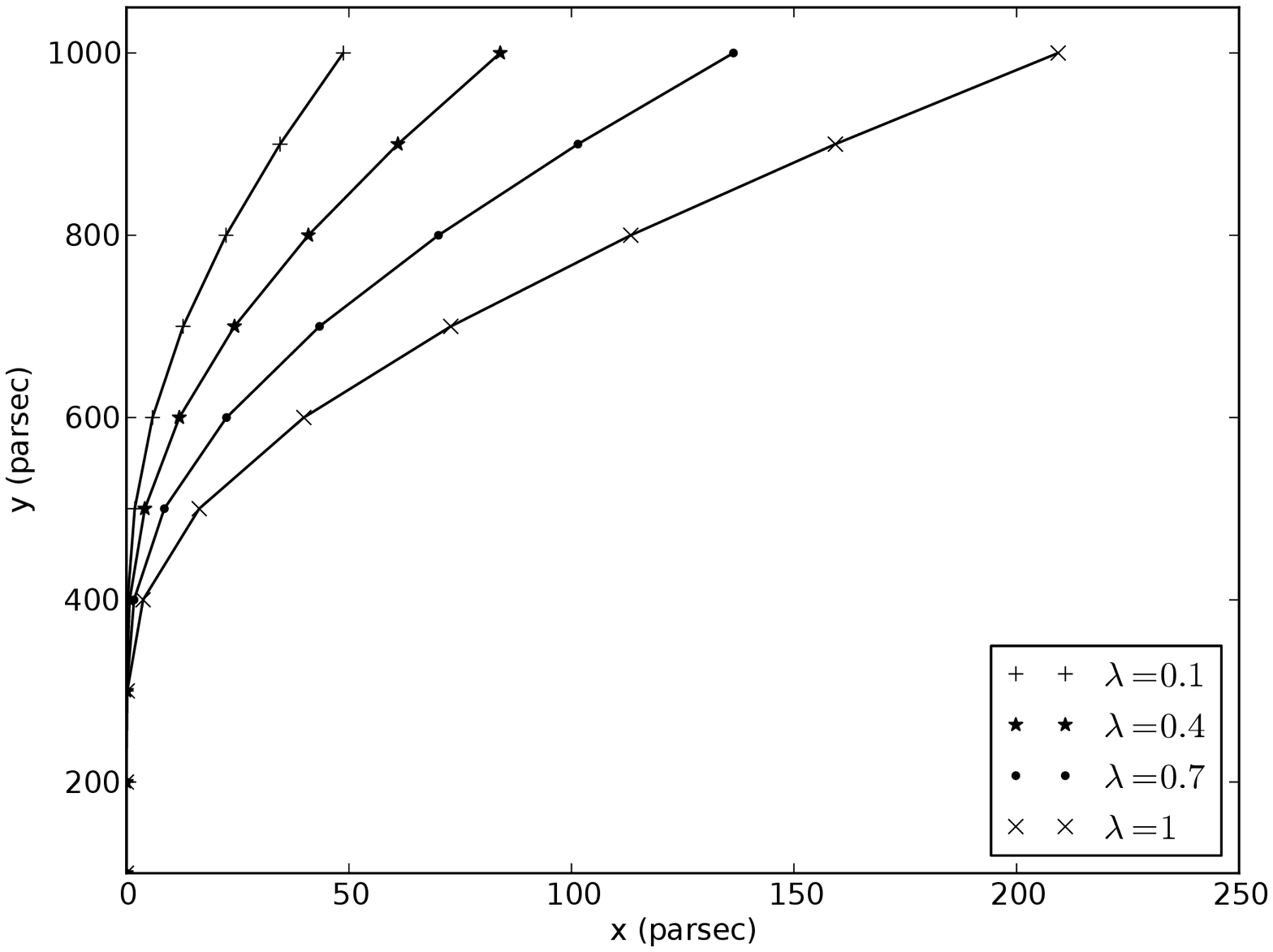} 
\caption{Jet paths with interaction constant $L=6.8284$ for all velocity coefficients $\lambda_i$. \label{f4}}
\end{figure}

\begin{figure}
\centering
\includegraphics[scale=0.5]{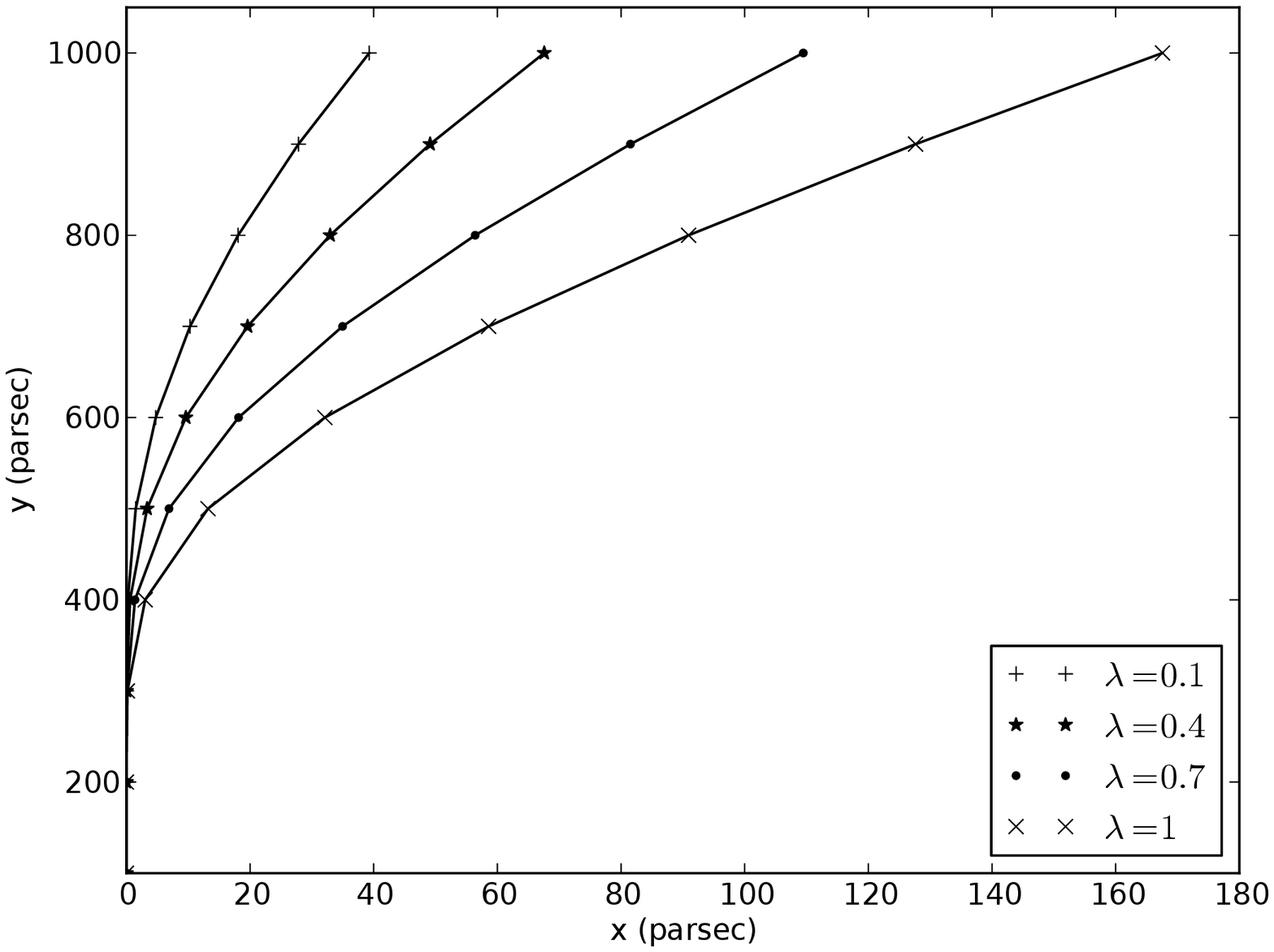} 
\caption{Jet paths with interaction constant $L=10.472$ for all velocity coefficients $\lambda_i$. \label{f5}}
\end{figure}

Numerical results for $\lambda_1 = 0.1$ and $\lambda_2 = 0.4$ are given in Table~\ref{tbl-1}; and for  $\lambda_3 = 0.7$, $\lambda_4 = 1.0$ in Table~\ref{tbl-2}. Grouping according to the interaction angle, Fig.~\ref{f2} shows the case $L_1 = 2.718$, Fig.~\ref{f3} the case $L_2 = 4$, Fig.~\ref{f4} the case $L_3 = 6.8482$, and Fig.~\ref{f5} deals with the case $L_4 = 10.472$. All  $X$ and $Y$ values are given in pc.

We can see how the velocity affects the shape of the jet, and how the interaction angle affects the final length of the jet. Moreover , in all cases computed, all values $X$ corresponding to $Y=100$~pc are of  order(s) below the accuracy of the computations, so considered to be zero. All $X$ values corresponding to $Y=200$~pc are of order(s) $10^{-4}$---$10^{-6}$. These values correspond to a diversion of approximately $10^{7}$---$10^{9}$~km; even thought it is very small compared to the scale of the jets, it cannot be neglected. We can conclude that the bending of the jet begins at a distance between 100~pc and 200~pc from the source.

Knowing the speed of the jet, and reducing observational data to scale, we can draw information and conclusion about the velocity variation and interaction angle.

\subsection{Kilo-parsec scale jets}
We consider the well-known jets of the structures 4C53.37 and 1159+583 in the galaxy clusters Abell 2220 and Abell 1446, respectively. The bending of the jets is now due to the relative movement of the galaxy through the ``intercluster medium'' (ICM).
Since, in this case, we have a much larger scale (hundreds of kilo-parsecs), we adopt a nonrelativistic hydrodynamic expression for the velocity given by the Euler equation 
\begin{equation}
\frac{\partial u}{\partial t} + (u\nabla)u = \, - \, \frac{\nabla P}{\rho} + g.
\end{equation}
Following a similar analysis as before, the expression for the velocity becomes (\citep{BB}, Eq.~(7); we use here the definitions and symbols of this investigation)
\begin{equation}
u = \left[
    \frac{6 \, R \, k \, T \, n_0 \, r}
         {\rho_t \, r_c^2 \, \left( 1 + r^2/r_c^2 \right)^{5/2}} + \, 
    \frac{2 \, R \, U^2_\mathrm{gal} \, m_p \, n_0}
         {\rho_t \, r_t \, \left( 1 + r^2/r_c^2 \right)^{3/2}}
    \right]^{1/2},
\label{uform}
\end{equation}
Burns and Balonek (\citep{BB}, Eq.~(9)) have managed to reduce Eq.~(\ref{uform}) to the form
\begin{equation}
u_j = U_\mathrm{gal} \, \left( \frac{n_\mathrm{ICM}}{n_j} \, 
                        \frac{2 \, R}{h} \right)^{1/2},
\end{equation}
where $u_j$ and $n_j$ are the velocity and density of the jet, $U_\mathrm{gal}$ the velocity of the host galaxy, $n_\mathrm{ICM}$ the mean intercluster density, $R$ the radius of curvature, and $h$ the distance from the source at which the interaction with the pressure gradient begins (scale height). For the case of ICM with high density, $h$ is equal to the radius of the galactic center; for the case of ICM with low density, on the other hand, $h$ coincides with the length of the jet.

\begin{figure}
\centering
\includegraphics[scale=0.5]{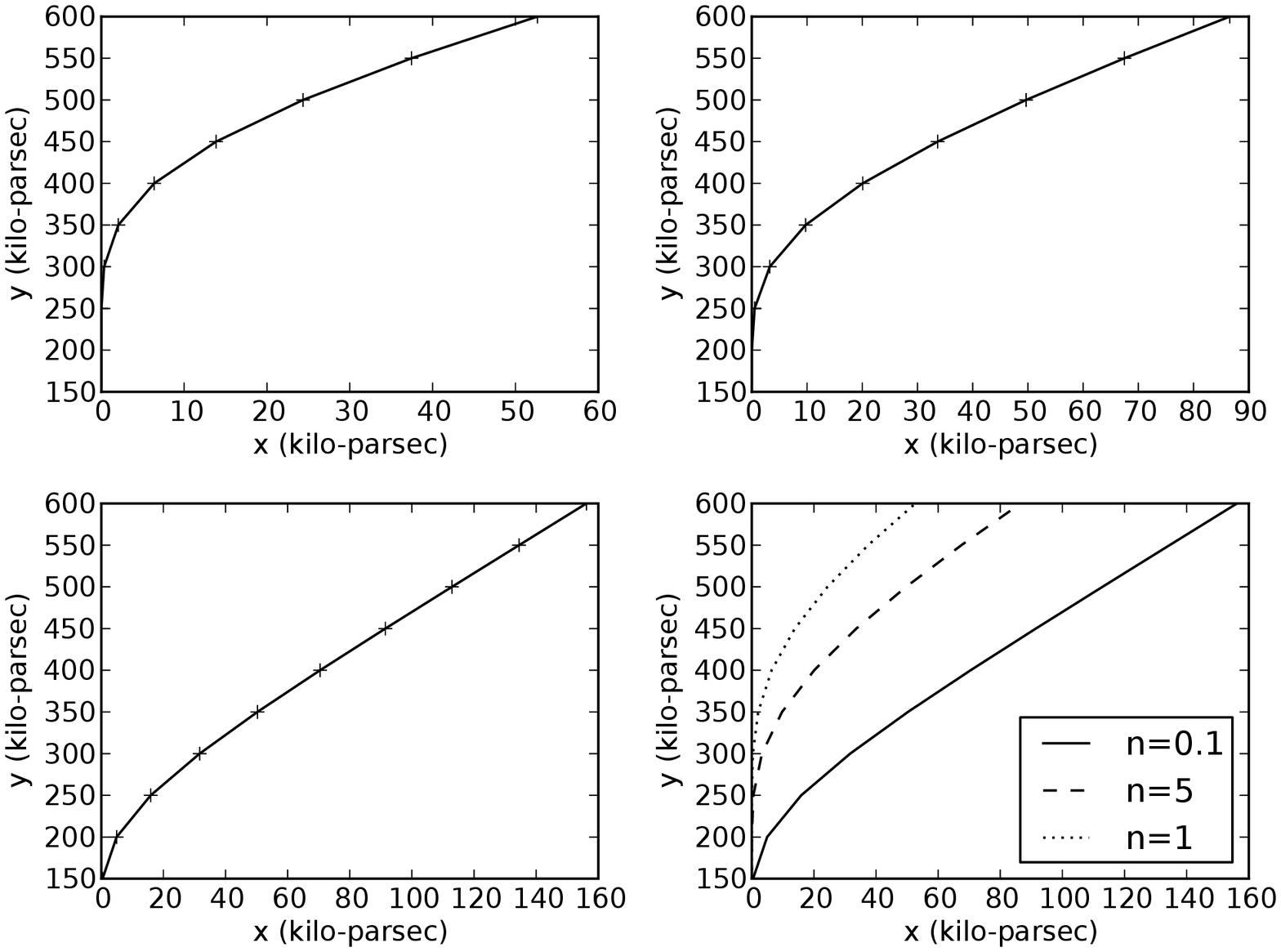}  
\caption{Jet paths for three density ratios. Top left: $n=1$. Top right: $n=0.5$. Bottom left: $n=0.1$. Bottom right: all $n$~values. \label{f6}}
\end{figure}

\begin{figure}
\centering
\includegraphics[scale=0.5]{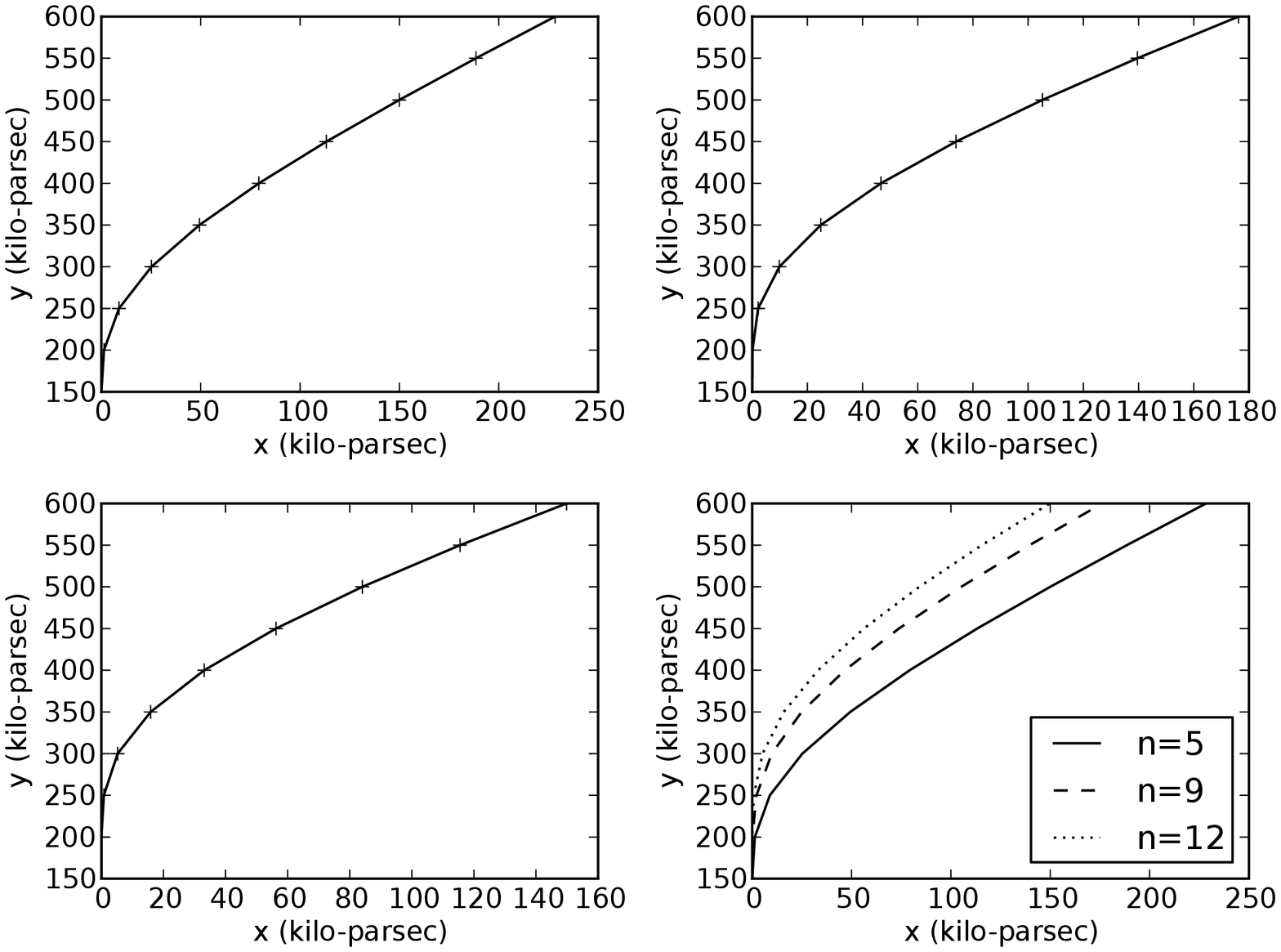}  
\caption{Jet paths for three density ratios. Top left: $n=5$. Top right: $n=9$. Bottom left: $n=12$. Bottom right: all $n$~values. \label{f7}}
\end{figure}

Using parameters given in \citep{BB} (Secs.~III, IV),  we plot three different values of the density ratio $n = n_{ICM}/n_j$ for each jet. Results for 4C53.37 and 1159+583 are shown in Table~\ref{tbl-3} and in Figs.~\ref{f6} and \ref{f7}, respectively.

The observational values for $n$ in \citep{BB} are $n = 0.5$ for 4C53.37, and $n = 12$ for 1159+583.
Figs.~\ref{f6} and \ref{f7} show how the shape of the jet does change with different intercluster densities. For the case of 4C53.37 there is a formidable variation in the shape of the curves between density ratios, whereas for the case of 1159+583 the differences are small. Comparing our results with observational data given by \citep{BO} and \citep{BOR} for 4C53.37 and 1159+583, respectively, and taking into account only the main jet, even though the jet in 4C53.37 seems to have lost its shape, if we imagine the path it would have followed, both jets seem to be in agreement with our computations.

\section{Conclusions}
Approximating a two dimensional kinematical model by some simplified simulations, we have studied the velocity profiles of large scale jets by assuming that their speed is not constant. We have seen that, when the velocity of the jet decreases linearly, the angle of interaction affects the length of the jet; and the bending begins at distances over 100~pc from the source. Regarding kilo-parsec scale jets, we can say that, morphologically, our simplified numerical simulations give results resembling relevant observations.

\begin{table}
\begin{center}
\caption{Calculated values with velocity coefficients $\lambda=0.1$  and $\lambda=0.4$ for all interaction constants. $X$ and $Y$ are given in pc. \label{tbl-1}}
\begin{tabular}{crrrrrrrr}
\hline\hline
 $X_{L_1}$ &$X_{L_2}$ & $X_{L_3}$ &$X_{L_4}$ & $X_{L_1}$  &$X_{L_2}$ & $X_{L_3}$  & $X_{L_4}$& Y  \\
\hline
  $10^{-6}$  & 8$\times10^{-7}$  & 6$\times10^{-7}$  & 5$\times10^{-7}$ & 5$\times10^{-6}$  & 4$\times10^{-6}$  & 3$\times10^{-6}$  & 2$\times10^{-6}$ & 200  \\
  0.012  & 0.01  & 0.007  & 0.006 & 0.039  & 0.032  & 0.025 & 0.02 &300 \\
  0.46  & 0.379  & 0.290  & 0.234 &  1.165  & 0.96  & 0.735  & 0.593 & 400 \\
  2.983  & 2.459  & 1.881  & 1.519 &  6.590  & 5.43  & 4.154  & 3.354 & 500 \\
  9.339  & 7.695  & 5.886  & 4.752 & 18.924  & 15.578  & 11.909  & 9.611 & 600 \\
  20.261  & 16.681  & 12.753  & 10.293 & 38.739  & 31.84  & 24.307  & 19.603 & 700 \\ 
  35.655  & 29.326  & 22.402  & 18.073 & 65.413  & 53.661  & 40.099  & 32.958 & 800  \\
  55.059  & 45.234  & 34.520  & 27.836 & 97.950  & 80.191  & 61.015  & 49.128 & 900 \\ 
  77.916  & 63.935  & 48.741  & 39.283 & 135.343  & 110.588  & 84.002  & 67.583 & 1000 \\  
\hline
\end{tabular}
\end{center}
\end{table}

\begin{table}
\begin{center}
\caption{Calculated values with velocity coefficients $\lambda=0.7$  and $\lambda=1$ for all interaction constants. Details as in Table~\ref{tbl-1}. \label{tbl-2}}
\begin{tabular}{crrrrrrrr}
\hline\hline
$X_{L_1}$ &$X_{L_2}$ & $X_{L_3}$ &$X_{L_4}$ & $X_{L_1}$  &$X_{L_2}$ & $X_{L_3}$  & $X_{L_4}$& Y  \\
\hline
  2.8$\times10^{-5}$  & 2.3$\times10^{-5}$  & 1.7$\times10^{-5}$  & 1.4$\times10^{-5}$ & 1.2$\times10^{-4}$  & $\times10^{-4}$  & 8$\times10^{-5}$  & 6$\times10^{-5}$ & 200 \\
  0.11  & 0.098  & 0.075  & 0.06 &0.326  & 0.269  & 0.206  & 0.166 &300 \\
  2.73  & 2.256  & 1.726  & 1.394 & 5.974  & 4.922  & 3.766  & 3.04 & 400 \\
  13.559  & 11.165  & 8.537  & 6.89 &26.086  & 21.445  & 16.374  & 13.207 & 500  \\
  35.89  & 29.488  & 22.506  & 18.148 & 63.989  & 52.396  & 39.871  & 32.106 & 600   \\
  69.59  & 57.017  & 43.408 & 34.961 & 117.964  & 96.157  & 72.893  & 58.599 & 700  \\
  113.080  & 92.356  & 70.127  & 56.411 & 184.833  & 150.027  & 113.330  & 90.939 & 800\\
  164.509  & 133.962  & 101.468  & 81.524 &  261.536  & 211.517  & 159.302  & 127.648 & 900 \\ 
  222.235  & 180.489  & 136.406  & 109.48 & 345.606  & 278.676 & 209.367  & 167.571 & 1000 \\
\hline
\end{tabular}
\end{center}
\end{table}

\begin{table}
\begin{center}
\caption{Calculated values with three different density ratios. For 4C53.37: $n_1=0.1$, $n_2=0.5$, $n_3=1$. For 1159+583: $n_4=5$, $n_5=9$, $n_6=12$. Details as in Table~\ref{tbl-1}. \label{tbl-3}}
\begin{tabular}{crrrrrrrrrrrr}
\hline\hline
  $X_{n_1}$ &$X_{n_2}$ & $X_{n_3}$ &$X_{n_4}$ & $X_{n_5}$  &$X_{n_6}$ & Y  \\
\hline
 0.408  & 1.3$\times10^{-6}$  & 0 & 0.016  & 8$\times10^{-5}$   & $10^{-6}$ & 150\\
  4.933  & 0.013  & $10^{-5}$ &  1.341  & 0.116  &  0.02 & 200 \\
  15.907  & 0.513  & 0.014 & 8.893  & 2.161  & 0.813  & 250 \\
  31.672  & 3.259  & 0.342  & 25.256  & 9.862  & 5.245 & 300 \\
  50.274  & 9.743  & 2.059  &  49.364  & 24.876  & 15.875 & 350 \\
  70.445  & 20.087  & 6.387 & 79.265  & 46.633  & 33.135 & 400 \\
  91.460  & 33.656  & 13.858  & 113.227  & 73.832  & 56.233 & 450 \\
  112.886  & 49.679  & 24.357  &  149.956  & 105.178  & 84.056 & 500  \\
  134.498  & 67.485  & 37.457  &  188.549  & 139.585  & 115.551 & 550 \\ 
  156.178  & 86.557  & 52.663  & 228.361  & 176.221  & 149.828 & 600  \\
\hline
\end{tabular}
\end{center}
\end{table}

\clearpage

\end{document}